\documentclass{jfm_old}
\usepackage{graphicx}
\usepackage{epstopdf,epsfig}
\usepackage{natbib}
\usepackage{hyperref}
\usepackage{natbib}
\usepackage{amsmath,mathtools}
\usepackage{graphicx}
\usepackage{adjustbox}
\hypersetup{
    colorlinks = true,
    urlcolor   = blue,
    citecolor  = black, 
}
\usepackage{color}
\usepackage{bm}
\usepackage{subcaption} 

\usepackage{tikz}
\usetikzlibrary{arrows.meta}

\newcommand{\RomanNumeralCaps}[1]
\linenumbers

\def\beq{\begin{equation}}
\def\eeq{\end{equation}}
\newcommand{\per}{\, .}
\newcommand{\com}{\, ,}

\newcommand{\act}{\mathcal{N}}
\newcommand{\actcirc}{\mathcal{N}^\circ}
\newcommand{\actst}{\mathcal{N}_{\star}}

\newcommand{\Cc}{\mathcal{C}}
\newcommand{\Uc}{\mathcal{U}}
\newcommand{\Vc}{\mathcal{V}}

\newcommand{\defn}{\overset{\textrm{def}}{=}}

\renewcommand{\p}{\partial}
\newcommand{\pal}{\partial_{\alpha}}
\newcommand{\pbe}{\partial_{\beta}}
\newcommand{\pt}{\partial_t}

\newcommand{\bcU}{\boldsymbol{\mathcal{U}}}

\newcommand{\ee}{\mathrm{e}}
\newcommand{\dd}{\mathrm{d}}
\newcommand{\ii}{\mathrm{i}}

\newcommand{\ka}{\kappa}

\newcommand{\Om}{\Omega}
\newcommand{\sig}{\sigma}
\newcommand{\al}{\alpha}
\newcommand{\bet}{\beta}
\newcommand{\bzetaa}{\boldsymbol{\zeta}^{\mathrm{a}}}
\newcommand{\bupsa}{\boldsymbol{\upsilon}^{\mathrm{a}}}

\newcommand{\bcirc}{{b}^\circ}

\newcommand{\bh}{\boldsymbol{h}}
\newcommand{\bx}{\boldsymbol{x}}

\newcommand{\bell}{\boldsymbol{\ell}}
\newcommand{\bk}{\boldsymbol{k}}
\newcommand{\bff}{\boldsymbol{f}}
\newcommand{\bxi}{\boldsymbol{\xi}}
\newcommand{\bu}{\boldsymbol{u}}
\newcommand{\bucirc}{\bu^\circ}
\newcommand{\ucirc}{u^\circ}

\newcommand{\bJ}{\boldsymbol{J}}

\newcommand{\cg}{\boldsymbol{\mathcal{C}}}
\newcommand{\cC}{\mathcal{C}}
\newcommand{\cG}{\mathcal{G}}

\newcommand{\grad}{\bnabla}
\newcommand{\bnablak}{\bnabla_{\!\bk}}
\newcommand{\bnablax}{\bnabla_{\!\bx}}
\newcommand{\curl}{\bnabla \!\times\!}
\newcommand{\diver}{\bnabla \bcdot}
\newcommand{\cross}{\times}
\newcommand{\eye}{\boldsymbol{\hat x}}

\newcommand{\kay}{\boldsymbol{\hat z}}
\def\bq{\boldsymbol{q}}
\def\ba{\boldsymbol{a}}

\def\bphi{\boldsymbol{\varphi}}
\def\bxi{\boldsymbol{\xi}}
\renewcommand\Re{\mathrm{Re}\,}
\def\balpha{\bm{\alpha}}
\def\bK{\bm{K}}

\newcommand{\mbk}{|\bk|}

\newcommand{\half}{\tfrac{1}{2}}

\newcommand{\Ewv}{\mathcal{E}_{\mathrm{w}}}

\newcommand{\buS}{\boldsymbol{u}^\mathrm{\scriptscriptstyle{S}}}

\newcommand{\buL}{\boldsymbol{u}^\mathrm{\scriptscriptstyle{L}}}

\newcommand{\uL}{u^\mathrm{\scriptscriptstyle{L}}}
\newcommand{\vL}{v^\mathrm{\scriptscriptstyle{L}}}
\newcommand{\wL}{w^\mathrm{\scriptscriptstyle{L}}}

\newcommand{\buE}{\boldsymbol{u}^\mathrm{\scriptscriptstyle{E}}}

\newcommand{\pse}{\boldsymbol{\mathsf{p}}}
\newcommand{\psest}{\boldsymbol{\mathsf{p}}_{\star}}

\newcommand{\ps}{\mathsf{p}}
\newcommand{\psst}{\mathsf{p}_{\star}}

\newcommand\JVcom[1]{\textcolor{cyan}{[[JV: #1]]}}
\newcommand\JV[1]{\textcolor{grey}{#1}}

\newcommand\ch[1]{#1}

\title{A consistent phase-averaged model of the interactions between surface gravity waves and currents}

\author{Jacques Vanneste\aff{1}
  \and
  William R. Young\aff{2}}

\affiliation{\aff{1}School of Mathematics and Maxwell Institute for Mathematical Sciences, University of Edinburgh, EH9 3FD, UK
 \aff{2} Scripps Institution of Oceanography, University of California at San Diego, La Jolla CA 92093-0213, USA
 }

\begin{document}
\maketitle

\begin{abstract}
We formulate a model of the two-way interactions between surface gravity waves and ocean currents. The model couples the transport of wave action in the four-dimensional (horizontal) position--wavevector phase space with the Craik--Leibovich system for the currents. Coupling is via the Doppler shift in the dispersion relation governing action transport, and wave pseudomomentum in the Craik--Leibovich system. The velocity in the Doppler shift is a vertical integral of the Lagrangian mean velocity of the currents, with a weight that is consistent with the vertical structure of the pseudomomentum. This consistency ensures conservation of momentum and energy in the coupled wave--current system.

The conservation properties of the wave--current model stem from an underlying variational structure. We derive this structure from that of the  rotating Euler equations for an incompressible fluid with free surface by introducing a Lagrangian wave--mean decomposition, making simplifying approximations, and Whitham averaging. \ch{The wave--mean decomposition is constructed such that the Lagrangian mean velocity is non-divergent and satisfies a rigid-lid boundary condition.}

We apply the wave--current model to the problem  of generation of inertial oscillations by surface waves originally considered by Hasselmann.

\end{abstract}

%

\section{Introduction \label{intro}}

Two-way interactions between ocean currents and high-frequency surface gravity waves dominate sea-surface dynamics. The  slow, phase-averaged evolution of the wave field  can be  modelled with the wave-action transport equation, \eqref{waveAct} below, \ch{governing dynamics in position--wavenumber phase space}. The action transport equation captures the effect of  currents on  waves, namely advection and refraction, through a Doppler-shift term in the dispersion relation. On the other hand,  the effect of waves on currents via the Stokes--Coriolis force \citep{Hassel1970,Huang1979} and the vortex force  is modelled by the Craik--Leibovich equation \citep{CL1976,leibovich1980} and its extensions \citep{ardhuin2008explicit,MRL2004,mcwilliams2022quasi}. Both forces depend on wave action through the Stokes velocity $\buS$. Doppler-shift and wave-averaged forces are the non-dissipative wave--mean-flow interactions by which energy and momentum are exchanged between waves and currents.

\ch{Although this exchange can proceed in both directions, a popular simplification considers the wave field for prescribed currents. This simplification underpins recent  work on the action transport equation \citep{villasboas2020wave,Boas2020,wang2023scattering,wang2025scattering,resseguier2024wave,HellChapron24} and the vast literature based on ray tracing \citep[e.g.][]{white1998chance,gallet2014refraction}.
 Likewise many studies solve the Craik--Leibovich equation \eqref{CL2} with  prescribed Stokes velocity $\buS(z)$ \citep[e.g.][]{skyllingstad1995ocean,mcwilliams1997langmuir,sesh2019surface,WagnerChini2021,wagner2023transition}.  Prescription of either waves or currents  severs the wave--mean-flow interaction problem into non-communicating halves.
 }

 Coupled  models  have  emerged in the last decade.  Driven by the demands of operational oceanography, established wave models based on the action transport equation are paired with  ocean circulation models \citep[e.g.][]{breivik2015surface,clementi2017,couvelard2020,sun2023,porcile2024}. This a posteriori coupling does not treat the wave action and Craik--Leibovich equations in a consistent manner. As a result, the conservation laws of energy and momentum are damaged.

Here we  formulate a  consistent  model describing the joint evolution of waves and currents (\S\ref{dcore}). Main elements of this framework are  familiar: currents enter the action transport equation via the Doppler shift $\bk \bcdot \bcU$ in the dispersion relation. But it is not obvious how the Doppler velocity $\bcU$ relates to the current in the Craik--Leibovich system. 
We show that taking $\bcU$ to be a suitable (wavenumber-dependent) weighted vertical average of the Lagrangian mean velocity $\buL$ leads to a coupled model conserving energy and momentum (\S\ref{seccons}).

The coupled model is consistent and enjoys conservation properties because of an underlying variational formulation. We obtain this formulation from a variational form of the rotating Euler equation with free surface by making a series of approximations: wave--mean decomposition, linear-wave ansatz  and Whitham averaging (\S\ref{varsec}).
This applies to surface gravity waves the approach pioneered by \citet{Dewar1970} and \citet{bret71} for hydromagnetic and acoustic waves. \ch{The approach was subsequently used by \citet{grimshaw1984wave}, \citet{gjajaholm} and \citet{salm16}  for internal and Poincar\'e waves, and by \citet{salmon2020more} and \citet{pizzo2021particle} for surface waves interacting with a two-dimensional quasigeostrophic flow \citep[see also][]{xie2015generalised,holm2023lagrangian}. Our treatment follows most closely \citeauthor{Dewar1970}'s (\citeyear{Dewar1970}) in relying on an Eulerian variational formulation and allowing for a broad wave spectrum. We exploit arbitrariness in the separation between mean-current and waves to constrain the Lagrangian mean velocity to be divergence free, following \citet{soward2010hybrid}, and the mean free surface to be flat.
A distinguishing feature of our model compared with previous work is that the Doppler velocity $\bcU$ is a weighted vertical integral of $\buL$ rather than $\buL$ itself because the sea surface is a waveguide.}

The exact conservation of energy and momentum in the absence of forcing and dissipation  for the coupled model of \S\ref{dcore} is key to understanding  wave--current interactions. Without a conserved energy it is impossible to decide on the `best' definition of mean kinetic energy, whether the Lagrangian mean kinetic energy $\frac{1}{2} |\buL|^2$ or its Eulerian counterpart $\tfrac{1}{2} |\buE|^2$, neither of which are conserved according to the Craik--Leibovich equation alone. Without  a consistently coupled wave action equation, the non-conservative terms in the energy equation of the currents can always be attributed to energy exchanges with the waves. 
 The consistent wave--current model in \S\ref{dcore} answers questions  that are inherently ambiguous when examined with either  prescribed Stokes velocity in the Craik--Leibovich equation or prescribed currents in the action equation. For example, with a complete understanding of energy conservation, we shown in \S\ref{seccons} and \S\ref{sec:hasselmann}  that the conserved energy involves the Lagrangian kinetic energy $ |\buL|^2/2$ rather than its Eulerian counterpart $ |\buE|^2/2$.

As a simple application,  we revisit the generation of inertial oscillations by surface waves considered by \citet{Hassel1970} (\S\ref{sec:hasselmann}). 
We replace \citeauthor{Hassel1970}'s prescribed Stokes velocity $\buS$ by a prescribed forcing of the action transport equation. Prescribed forcing of the action  represents the effect of wind more directly than prescription of $\buS$ in the Craik--Leibovich equation. The current dynamics obtained with the wave--current model is identical to that of \citeauthor{Hassel1970} but the energetics of the process is now clear: the energy of the inertial oscillations is not drawn from the wave energy but instead requires additional wind work.



\section{A consistent wave--current model}\label{dcore}

The coupled model summarised in this section -- referred to as `consistent wave--current model', CWCM for short -- follows from application of  Hamilton's principle to an approximate Lagrangian, given in \eqref{lagrangian3}. The derivation of this approximate Lagrangian, starting  from the exact Lagrangian for rotating incompressible flow in \eqref{lagrangian1},  is technically demanding. Instead it is economical to  first present the CWCM, then verify its conservation laws (\S\ref{seccons}).

The phase-averaged wave field is characterized by the wave action density $\act(\bx,\bk)$. Our notation  is that $\bk=(k,l)=(k_1,k_2)$ and $\bx=(x,y)=(x_1,x_2)$  are horizontal vectors; unless emphasis is required we do not  indicate  dependence  of various fields  on  time $t$. We indicate functions of the phase space variables $(\bx,\bk)$ using calligraphics, e.g. the action density  $\act(\bx,\bk)$ and also $\bcU(\bx,\bk)$ and $\cg(\bx,\bk)$ below. We make  exceptions for fields traditionally denoted by  Greek letters, e.g. the absolute and intrinsic frequencies $\Omega(\bx,\bk)$ and $\sigma(\bx,\bk)$ below. The action density $\act(\bx,\bk)$ is the phase-space energy density divided by intrinsic frequency $\sigma(\bx,\bk)$ and, in our convention, divided by the constant fluid density $\rho$.

The evolution of the phase-averaged wave field is determined by the action balance equation: 
\beq
\partial_t \act  + \bnablak \, \Om \bcdot \bnablax \act - \bnablax\, \Om \bcdot \bnablak  \, \act = \actcirc
\label{waveAct}
\eeq
\citep[e.g.][]{holthuijsen2010waves,janssen2004interaction,KomenBook}. $\actcirc$ on the right of \eqref{waveAct} denotes sources and sinks of wave action  and nonlinear interactions between waves. \ch{We consider water with depth $d(\bx)$ so  that} the wave frequency in  \eqref{waveAct} is
\beq
\Om(\bx,\bk) = \sig(\bx,\ka)+ \bk\bcdot \bcU(\bx,\ka)\com \label{omIntro}
\eeq
where 
\beq
\sig(\bx,\ka) =\sqrt{g \kappa \tanh\big(\kappa \, d(\bx)\big)}, \quad \textrm{with} \quad \ka = \mbk = \sqrt{k^2+l^2},
\label{intrinsic}
\eeq 
is the intrinsic frequency. In \eqref{omIntro} the Doppler shift $\bk\bcdot \bcU$ is based on  the `Doppler velocity' 
\beq
\bcU(\bx,\ka) \defn \int^0_{-d(\bx)} \!\!\!\!\!\! \buL(\bx,z) Q(\bx,z,\kappa) \, \dd z,
\label{Udop}
\eeq
where $\buL(\bx,z)$ is the three-dimensional Lagrangian mean velocity and 
\beq
Q(\bx,z,\kappa) \defn  \frac{2 \kappa \cosh(2 \kappa (z+d(\bx)))}{\sinh(2 \kappa d(\bx))}
\label{Q}
\eeq
is such that
\beq
\int_{-d(\bx)}^0  \!\!\!\! Q(\bx, z, \ka) \, \dd z = 1\per
\label{Qnorm}
\eeq
Because $\bk$ is a horizontal vector, the vertical component of $\bcU$ plays no role in \eqref{waveAct}--\eqref{omIntro} nor in the sequel.


\ch{In the  derivation of the CWCM in section \ref{varsec}  the `vertical averaging kernel', $Q(\bx,z,\ka)$ in \eqref{Q}, is produced by inserting the  linear-wave ansatz \eqref{vstru} into a Lagrangian and Whitham-averaging. The resulting} dispersion relation \eqref{omIntro} with \eqref{Udop}--\eqref{Q} is also  obtained by \citet{kirby1989surface} as an approximation to the dispersion relation of finite-depth waves propagating in a vertically sheared current.
In deep water, 
\beq
Q(\bx,z,\kappa) = 2 \kappa \, \ee^{2 \ka z},
\eeq
as originally obtained by \citet{stewart1974HF}.

The dispersion relation  with \eqref{Udop} and \eqref{Q}  is used by \citet{ardhuin2008explicit} in their extension of the Craik--Leibovich system. In all these earlier instances, the Doppler velocity involves the Eulerian mean velocity $\buE$ instead of the Lagrangian mean velocity $\buL$ in \eqref{Udop}.
\ch{In contrast, the systematic analysis of small-amplitude single wavetrains in deep water using Lagrangian coordinates leads to the Doppler velocity in the form \eqref{Udop} with $d \to\infty$ \citep{abrashkin2017lagrange,abrashkin2018dynamics,pizzo2023role}.}

\ch{The validity of the dispersion relation \eqref{omIntro} requires the shear to be weak and is ensured provided that $|\partial_z \buL_\mathrm{h}| \ll \sigma$, with $\buL_\mathrm{h}$ the horizontal Lagrangian mean velocity \citep[cf.][]{banihashemi2019approximation}.}

The evolution of the Lagrangian mean current $\buL(\bx,z)=(\uL,\vL,\wL)$ is determined by  the  Craik--Leibovich system 
\begin{align}
\p_t (\buL - \pse)  +  \big(\bff + \bnabla \times (\buL -  \pse)  \big)\cross \buL+ \bnabla  \pi &=\bucirc,  \label{CL2}  \\
\diver \buL&=0,\label{divuL} 
\end{align}
where $\bnabla = (\bnablax,\partial_z)$ is the three-dimensional spatial gradient. In \eqref{CL2} $\bff$ is a constant vector;  we do not make the traditional approximation. Coupling  between the action conservation  \eqref{waveAct} and the Lagrangian mean velocity $\buL(\bx,z)$  is via  the pseudomomentum
\beq
\pse(\bx,z) \defn \iint  Q(\bx,z,\kappa)  \act(\bx,\bk) \bk \, \dd \bk\per
\label{pmom1}
\eeq
Pseudomomentum $\pse(\bx,z)$  is a horizontal vector. Integrating \eqref{pmom1} over $z$ and using \eqref{Qnorm} produces the important identity
\beq
\int_{-d(\bx)}^0 \!\!\! \pse(\bx,z) \, \dd z = \iint   \act(\bx,\bk) \bk \, \dd \bk\per
\label{pmomint}
\eeq

The Craik--Leibovich system is completed with the $z=0$  rigid-lid boundary condition on the Lagrangian mean vertical velocity
\beq
 \wL(x,y,0,t)=0\com
\label{rigidLid}
\eeq
and the no-normal-flow bottom boundary condition
\beq
\wL + \buL \bcdot \bnabla_{\bx} d = 0 \quad \textrm{at} \ \ z = - d (\bx).
\label{bottombc}
\eeq
The rigid-lid boundary condition ensures  that \eqref{CL2} and \eqref{divuL} does not have wave solutions, i.e.\ waves are filtered from the wave-averaged equations.

\begin{figure}
\begin{center}
\begin{tikzpicture}[scale=1]

\fill[blue!20] (0, 0.5) rectangle (6.0 , 3.0);
\fill[orange!20!white] (0,5) rectangle (6 , 7.5);
\node at (3,1.75) {$\buL$};
\node[right] at (0,4) {$\bm{\mathcal{U}} = \int \buL Q \, \dd z$};
\node[left] at (6,4) {$\pse = \int \act Q  \bk \, \dd \bk$};
\node at (3,6.25) {$\act$};

\draw[thick,blue] (1.3,3) -- (1.3,3.8);
\draw[-{Latex[open,scale=1.4]},thick,blue] (1.3,4.2) -- (1.3,5);
\draw[-{Latex[open,scale=1.4]},thick,orange]  (4.7,3.8) -- (4.7,3);
\draw[thick,orange]  (4.7,5) -- (4.7,4.2);

\draw[->, thick] (0.6, 1) -- (2.1, 1) node[below,yshift=-.2ex] {$\bm{x}$};
\draw[->, thick] (0.6, 1) -- (0.6, 2.5) node[left,xshift=-.3ex] {$z$};
\draw[->, thick] (0.6, 5.5) -- (2.1, 5.5) node[below,yshift=-.2ex] {$\bm{x}$};
\draw[->, thick] (0.6, 5.5) -- (0.6, 7) node[left,xshift=-.3ex] {$\bm{k}$};

\end{tikzpicture}
\caption{Schematic of the CWCM: the dynamical variables are the action density $\act(\bx,\bk)$ and the current Lagrangian mean velocity $\buL(\bx,z)$. They are coupled through the pseudomomentum $\pse(\bx,z)$ and the weighted horizontal velocity $\bcU(\bx,\ka)$ that appears in the surface wave dispersion relation.}
 \label{fig:scheme}
\end{center}
\end{figure}

The  system in \eqref{waveAct} through \eqref{bottombc} is the dynamical  core of the CWCM. Additional processes, such as forcing, dissipation and wave--wave interactions, can be added   by specification of the right hand sides, $\actcirc$ and $\bucirc$,  of  \eqref{waveAct} and \eqref{CL2}. The model combines the dynamics of the action in the four-dimensional $(\bx,\bk)$ phase space with Craik--Leibovich dynamics in the three-dimensional $(\bx,z)$ physical space. The two components of the model interact via the pseudomomentum $\pse(\bx,z)$ and the horizontal Doppler velocity $\bcU(\bx,\kappa)$, both of which involve  the weight function $Q(\bx,z,\kappa)$. This is illustrated in figure \ref{fig:scheme}. 


The CWCM uses the Lagrangian mean velocity $\buL$ and the pseudomomentum $\pse$;    the Stokes velocity $\buS$ and the Eulerian mean velocity $\buE$  do not appear.  For approximately irrotational waves, as assumed in  the derivation in  \S\ref{varsec},  $\buS \approx \pse$. Thus one can compare the CWCM with other formulations of the mean momentum equation that employ $\buS$ and  $\buE=\buL- \buS \approx \buL - \pse $  as main variables \citep[e.g.][]{CL1976,MRL2004,SuzukiFoxKemper}.

An unfamiliar feature of the CWCM  is that the Lagrangian mean $\buL$, depth-weighted in \eqref{Udop}  with the kernel $Q(\bx,z,\ka)$,   is  the Doppler velocity $\bcU(\bx,\ka)$ in the dispersion relation \eqref{omIntro}.  Because  $\bcU$ depends on  $\ka$   the group velocity is
\beq
\cg(\bx,\bk) = \bnablak \Om  =\cG (\bx,\bk) \bk + \bcU(\bx,\ka)\com
\label{cg0}
\eeq
where
\beq
\cG (\bx,\bk) \defn  \frac{1}{\ka} \frac{\p\sig}{\p \ka}+  \frac{\bk}{\ka}\bcdot \frac{\partial \bcU}{\partial \ka} \per \label{GaDef}
\eeq
\cite*{banihashemi2017approximation} and \cite{banihashemi2019approximation} have drawn attention to $\kappa$-dependence of  the current in the wave action conservation and the peculiar term involving $\partial \bcU/\partial \ka$ in $\cG$.


For reference we record alternative equivalent forms of the Craik--Leibovich equation \eqref{CL2}.   \cite{AndrewsMcIntyre1978} and \cite{leibovich1980} favour the form that emerges directly from the application of Hamilton's principle in  \S\ref{varsec},
\beq
(\p_t + \uL_{j} \p_j)(\uL_i - \ps_i) +  (\uL_j - \ps_j) \p_i\uL_{j}  + (\bff \cross \buL)_{i} +\p_i \hat \pi = \ucirc_{i}\per
\label{AML}
\eeq
In \eqref{AML} italic indices $i$ and $j$  run from $1$ to $3$, $\p_i$ is shorthand for $\p /\p x_i$ and 
\beq
\hat \pi \defn \pi  + \pse \bcdot \buL - |\buL|^2\per
\eeq

 Another equivalent form is 
\beq
\p_t \uL_i - \p_t \ps_i - \ps_{j} \p_i\uL_{j}   + \p_j(\uL_i \uL_j - \ps _{i}  \uL_{j})+ (\bff \cross \buL)_i + \p_i \tilde \pi = \ucirc_{i} \com
\label{momCons0}
\eeq
where
 \beq
 \tilde \pi = \pi + \pse \bcdot \buL - \half|\buL|^2\per
 \eeq
 The form in \eqref{momCons0} is useful in the discussion of momentum conservation in \S\ref{momCons}. \citet{SuzukiFoxKemper} present five alternative versions of the Craik--Leibovich equation with $\buE$ or $\buL$ as the primary velocity. All versions  follow from \eqref{CL2}, \eqref{AML} or \eqref{momCons0} on using $\buL - \pse = \buE$ and $\pse = \buS$.
 

\section{Conservation laws} \label{seccons}

The CWCM in  \S\ref{dcore} conserves action, mass,  energy, momentum, circulation and potential vorticity. According to \eqref{waveAct}, the action density $\act(\bx,\bk)$ is transported by the non-divergent phase-space velocity $(\bnablak \, \Om , - \bnablax \, \Om)$. \ch{The corresponding conservative form of \eqref{waveAct} is
\beq
\partial_t \act  + \bnablax \bcdot (\act \bnablak \, \Om  ) - \bnablak\bcdot (\act \bnablax\, \Om  )= \actcirc \per
\label{waveActCons}
\eeq
If $F(\cdot)$ is any function, multiplying \eqref{waveActCons} by the derivative $F'(\act(\bx,\bk))$  shows that
\beq
\iint F(\act(\bx,\bk)) \, \dd \bx \dd \bk
\label{Fint3}
\eeq
is conserved  when $\actcirc = 0$.  The conservative  form in \eqref{waveActCons} is also the most convenient for forming $\bk$-weighted moments of the action equation e.g. as below in \eqref{momCons4}.
}

\subsection{Mass conservation}

The Lagrangian average used in the formulation above, with a non-divergent $\buL$ in \eqref{divuL}, is not the GLM Lagrangian average of \cite{AndrewsMcIntyre1978}: in GLM  $\buL$ has non-zero divergence \citep[][]{McIntyre1988,ardhuin2008explicit}.   Instead we use the glm average of \cite{soward2010hybrid} which has a non-divergent Lagrangian mean velocity \cite[][]{vanneste2022}.

(Mass is conserved by GLM: with  non-zero $\diver \buL$ there is a Lagrangian density variable -- denoted $\tilde \rho$ in \cite{AndrewsMcIntyre1978} --  ensuring   mass conservation.  Our decision to use  glm, with exact $\diver \buL=0$, avoids the additional $\rho$-variable.)

\subsection{Momentum conservation} \label{momCons}

We start with the vertically integrated pseudomomentum equation obtained by forming    $\iint \eqref{waveActCons} \bk \, \dd \bk$:
\beq
\pt \iint  \act(\bx,\bk) k_{\al}\, \dd \bk +\pbe \iint \cC_{\bet} k_{\al}\act \dd \bk + \iint  \act \pal\Om \, \dd \bk = \iint k_{\al}\actcirc \, \dd \bk \per
\label{momCons4}
\eeq
In \eqref{momCons4} the Greek indices $\al$ and $\bet$  run from $1$ to $2$ \ch{and $\pal$ and $\pbe$ denote  derivatives with respect to $x_\al$  e.g. $\pal =\p/\p x_\al$. The group velocity, $\cC_\bet$ above,  is defined  in \eqref{cg0}. With pseudomomentum $\pse(\bx,z)$ defined in  \eqref{pmom1}, the first term in \eqref{momCons4} is $\pt$ applied to the total vertical integral of  $\pse(\bx,z)$. The third term in \eqref{momCons4}, involving $\pal \Om$, is transformed by switching the order of $\bk$- and $z$-integrals; see Appendix \ref{vardepth} for details.
These manipulations bring \eqref{momCons4} to the form
  \begin{align}
\pt \int \! \ps_\al \, \dd z   + \pbe \iint \cC_{\bet} k_{\al} \act \, \dd \bk +  \int \ps_{\beta}\pal \uL_{\beta}  \, \dd z   +  \iint \frac{\p \Om}{\p d} \act \dd \bk\,  \pal d  = \iint k_{\al}\actcirc \, \dd \bk\per
\label{momCons5}
\end{align}
 Above and hereafter we lighten notation by suppressing the limits $-d(\bx)$ and $0$ on vertical integrals.  In writing the derivative $\p \Om/\p d$ we abuse notation slightly by treating $\Omega$ as a function of $d$ rather than $\bx$ as in \eqref{omIntro}. The interpretation is clear from the explicit dependence of $\sigma$ and $\bcU$ (via $Q$) on $d$ in \eqref{intrinsic} and \eqref{Udop}--\eqref{Q}. We give an explicit expression for  $\p \Om/\p d$ in \eqref{dHell3} and \eqref{dHell7}. 
 }
 
 
 
Turning to the Craik--Leibovich equation, the vertical integral of the horizontal components of \eqref{momCons0} is
\begin{align}
\p_t \! \int \!\uL_{\al} \, \dd z -& \p_t \!\int \!\ps_\al \dd z - \!\int \ps_{\bet} \pal \uL_{\bet} \, \dd z +  \p_{\bet }\!\int \!\uL_\al \uL_\beta - \ps_\al \uL_\bet \, \dd z  \nonumber \\
&\quad + \Big(\int \!\bff \cross \buL \, \dd z \Big)_{\al}+ \p_{\al} \!\int \tilde \pi\, \dd z  - [ \tilde \pi]_{-d} \, \pal d =    \int \!\ucirc_\alpha  \, \dd z\com
\label{momCons2}
\end{align}
where $[\,\cdot\,]_{-d}$ denotes evaluation at $z=-d$, e.g. $ [ \tilde \pi]_{-d}  = \tilde \pi(\bx, -d(\bx))$.
Terms stemming from the commutation of $\partial_\beta$ with the vertical integral vanish on using the boundary conditions \eqref{rigidLid} and \eqref{bottombc}. The sole remaining boundary contribution is the topographic form stress $[ \tilde \pi]_{-d} \, \pal d$ in \eqref{momCons2}.

Adding  \eqref{momCons5} and \eqref{momCons2} cancels the terms
\beq
\p_t \int\!\! \ps_\al \, \dd z  +  \int \ps_{\beta}\pal\uL_{\beta}  \, \dd z \com
\eeq
leaving  
\begin{align}
\p_t &\!\int \!\uL_{\al} \, \dd z +  \p_{\bet }\left(\int \!\uL_\al \uL_\beta - \ps_\al \uL_\bet  \,\dd z  + \iint k_\al \cC_\bet  \act \dd \bk \right) + \Big(\int \!\bff \cross \buL\dd z \Big)_{\al} \nonumber \\
& + \p_{\al} \!\int \tilde \pi\, \dd z  +  \bigg( \iint \frac{\p \Om}{\p d} \act \dd \bk \,   -   [\tilde \pi]_{-d} \bigg) \pal d  =    \int \! \ucirc_\alpha  \dd z + \iint k_\al \actcirc \, \dd \bk\per
\label{momCons3}
\end{align}
The result
\beq
\int \ps_\al \uL_\beta \, \dd z = \iint k_\al\, \bcU_\bet \, \act \dd \bk\com 
\label{ref2}
\eeq
deduced from the definition of the Doppler velocity $\bcU$ in  \eqref{Udop} and the definition of pseudomomentum $\pse$ in \eqref{pmom1},  produces a further partial cancellation in \eqref{momCons3}:
\begin{align}
\iint k_\al \cC_\bet  \act \dd \bk - \int \ps_\al \uL_\bet \, \dd z  
&=\iint \cG k_\al k_\bet\, \act \, \dd \bk\com \label{pcancel}
\end{align}
where $\cG(\bx,\bk)$ is defined in \eqref{GaDef}.  Thus   \eqref{momCons3} becomes
\begin{align}
\p_t & \!\int \!\uL_{\al} \, \dd z +  \p_{\bet}\left(\int \!\uL_\al \uL_\bet \, \dd z + \iint \cG \, k_\al  k_\bet  \act \dd \bk \right) + \Big(\int \!\bff \cross \buL\dd z \Big)_{\al}\nonumber \\
& + \p_{\al} \!\int \tilde \pi\, \dd z +\bigg( \iint \frac{\p \Om}{\p d} \act \dd \bk \,   -  [ \tilde \pi]_{-d} \bigg) \pal d 
 =    \int \! \ucirc_\al  \dd z + \iint k_\al \actcirc \, \dd \bk\per
\label{momCons111}
\end{align}
This is momentum conservation in the absence of the Coriolis term, depth variation $\pal d$, and of forcing and dissipation on the right-hand side.


Although the pseudomomentum $\pse$ appears throughout the manipulations above, remarkable cancellations ensure that $\pse$ does not feature in the final form 
\eqref{momCons111}. This is a manifestation of \citeauthor{McIntyre1981myth}'s (\citeyear{McIntyre1981myth}) `wave momentum myth': the \ch{total} momentum density, namely
\beq
 \!\int \buL \, \dd z  \com
\label{totalmomentum}
\eeq
contains only the Lagrangian mean velocity and the direct effect of the waves in \eqref{momCons111} is flux of momentum  via  the radiation stress tensor
\beq
\iint \cG \, k_\al  k_\bet  \act \dd \bk\com
\label{radiationstress}
\eeq
where $\cG (\bx,\bk)$ is defined in \eqref{GaDef}.

\ch{As noted by \citet{Dewar1970} we can regard \eqref{totalmomentum} as the sum of the Lagrangian mean momentum of the current
\beq
 \int \buL - \pse \, \dd z
\eeq
and the wave pseudomomentum  $\int \pse \, \dd z$ (often termed  wave momentum). This is consistent with a geometric interpretation of the Lagrangian mean momentum \citep{GilbertVanneste2018,gilbert2025geometric}. The total momentum \eqref{totalmomentum} is the conserved quantity associated with translational symmetry of the system.}

In this section, for simplicity, we have focused  on the depth-integrated momentum. By forming  $\iint \eqref{waveAct} Q \bk \, \dd \bk$ and combining with the Craik--Leibovich equation \eqref{momCons0}, it is possible to derive an equation for the total momentum $\buL$ at each $z$. This equation reduces to \eqref{momCons111} when integrated with respect to $z$. It provides an alternative to \eqref{momCons0} in which wave forcing appears through radiation stresses with vertical as well as horizontal components.
%

\subsection{Energy conservation \label{energyCons}}

Following \cite{bretherton1968}, we expect that the wave-energy density is
\beq
\Ewv(\bx,\bk)=\sig(\bx,\ka) \, \act(\bx,\bk)\com
 \label{Ewv}
\eeq
where $\sig(\bx,\ka)$ is the intrinsic frequency in \eqref{intrinsic}.
We also expect   that the total energy density,
\beq
E(\bx) = \int \half |\buL|^2 \, \dd z + \iint \Ewv(\bx,\bk) \, \dd \bk\com
 \label{Econs3}
 \eeq
 satisfies a physical space conservation law
 \beq
 \partial_t E + \bnablax \!\bcdot \!\bJ = \, {E}^\circ\per
 \label{Econs5}
 \eeq
 The horizontal vector  $\bJ(\bx)$ above  is the energy flux.

To determine $\bJ$ and ${E}^\circ$ in \eqref{Econs5}, start with  the depth integral of $\buL \bcdot \eqref{CL2}$ and use the vertical boundary conditions  \eqref{rigidLid} and \eqref{bottombc} to find
\beq
\p_t \int  \half |\buL|^2 \, \dd z +\bnablax \bcdot \int \buL \pi \, \dd z= \int  \buL\bcdot \pt \pse\, \dd z + \int \buL\bcdot \bucirc \dd z \per \label{energ17}
\eeq
Equation \eqref{energ17} is the energy equation for the currents.

\ch{Although  the wave energy $\Ewv$ in \eqref{Ewv} is related to action density $\act(\bx,\bk)$ by  multiplication with the intrinsic frequency $\sigma(\bx,\bk)$, the wave energy conservation law is obtained with  $\iint \eqref{waveActCons} \Om \, \dd \bk$ i.e. multiplication of the action conservation equation by the total frequency $\Om(\bx,\bk)=\sigma(\bx,\ka) + \bk\bcdot \bcU(\bx,\ka)$. } This produces
\beq
\pt  \iint \Ewv \, \dd \bk + \iint (\bk \bcdot \bcU) \pt \act \, \dd \bk + \bnablax \bcdot  \iint \cg \Omega \act \dd \bk = \iint \Om \actcirc \dd \bk \com \label{energ19}
\eeq
where $\cg$ is the group velocity in \eqref{cg0}. The key identity 
\beq
\iint ( \bk \bcdot \bcU ) \, \pt \act \, \dd \bk = \int \!\buL \! \bcdot \! \iint Q \bk \, \pt \act \, \dd \bk \, \dd z = \int \!\buL \!\bcdot \pt \pse \, \dd z
\eeq 
then puts \eqref{energ19} in the form
\beq
\pt \iint \Ewv\, \dd \bk + \int \buL \bcdot \pt \pse \, \dd z + \bnablax \bcdot  \iint \cg \Omega \act \dd \bk = \iint \Om \actcirc \dd \bk \per \label{energ23}
\eeq

Summing \eqref{energ17} and  \eqref{energ23}  eliminates the energy-exchange term 
\beq
\int  \buL\bcdot \pt \pse\, \dd z
\eeq
and produces energy conservation in  \eqref{Econs5} with 
\beq
\bJ = \int \buL \pi \, \dd z +  \iint \cg \Om \act \, \dd \bk\com
\label{energ29}
 \eeq 
and  
\beq
{E}^\circ = \iint \Om \actcirc \, \dd \bk + \int \buL \bcdot \bucirc \, \dd z \per \label{energ31}
\eeq
The right of \eqref{energ23},
\beq
\iint \Om \actcirc \dd \bk = \iint \sigma  \actcirc \dd \bk + \iint  \bk \bcdot \bcU \, \actcirc \dd \bk\com
\eeq
involves a non-obvious contribution of the Doppler shift to ${E}^\circ$. We return to this term  in \S\ref{sec:hasselmann}.

Using the dispersion relation \eqref{omIntro}, the final term in \eqref{energ29} can be split into physically distinct contributions
\beq
\iint \cg \Om \act \, \dd \bk = \iint \cg  \Ewv \, \dd \bk + \iint \cG \bk \bk\bcdot \bcU \act \dd \bk + \iint \bcU \bcU \bcdot \bk \act \, \dd \bk  \per
 \label{energ37}
\eeq
The first term on the right of \eqref{energ37} is group-velocity transport of the wave energy $\Ewv$. The second term is the wave radiation stress \eqref{radiationstress} acting on $\bcU$. The third term is the momentum flux associated with $\bcU$ acting on the wave pseudomomentum. 

The mean kinetic energy in the conserved energy \eqref{Econs3} is the Lagrangian kinetic energy $\half |\buL|^2$ rather than the Eulerian kinetic energy $\half |\buE|^2$. For the CWCM, this is natural: $\buL$ is the primary dynamical variable and $\buE$ does not appear explicitly. Moreover, the form of the conserved energy  is linked  to the variational formulation of the CWCM in \S\ref{varsec} via Noether's theorem. 

 \citet{WagnerChini2021} and \cite{czeschel2023energy} contrast the simplicity of the Lagrangian kinetic energy budget for the (uncoupled) Craik--Leibovich system with the complexity of the Eulerian kinetic energy budget. The CWCM explains this simplicity by the privileged role of the Lagrangian kinetic energy as one of the two terms in the conserved total energy \eqref{Econs3}.
 In \S \ref{sec:hasselmann} we return to the primacy of $\half |\buL|^2$ over $\half |\buE|^2$  in a discussion of wave-forced inertial oscillations. In this basic example the CWCM is forced by prescribing $\actcirc$ in the action transport equation \eqref{waveAct} and all terms in the energy balance can be calculated explicitly.

\subsection{Circulation and potential vorticity}

Material conservation of circulation and potential vorticity is a property of the Craik--Leibovich equation \eqref{CL2},  i.e.\ these  laws are independent of the wave action equation \eqref{waveAct}.  We state the  results and defer to \cite{AndrewsMcIntyre1978} or \citet{holm1996} for proof.  

Introduce the `absolute Lagrangian mean momentum'
\beq
\bupsa =  \bh +  \buL -  \pse  \com
\label{buspsa}
\eeq
where
\beq
\bh(\bx_3) = \tfrac{1}{2} \bff \times \bx_3
\label{hdef}
\eeq   
with $\bx_3 = (\bx,z)$. A key property of $\bh$ is  that $\curl  \bh = \bff$. The corresponding absolute vorticity is
\beq
\bzetaa = \curl \bupsa=\bff + \bnabla \times (\buL -  \pse)\per
\eeq
Then conservation of circulation is
\beq
\frac{\dd}{\dd t} \oint_{\Cc} \bupsa \bcdot  \dd \bell=0 \com
\eeq
where $\Cc$ is a material circuit and we assume $\bucirc =  0$. 

As noted by \citet{AndrewsMcIntyre1978}, for $\bff = 0$, conservation of circulation implies that $\bnabla \times (\buL -  \pse) = 0$ for all times if this holds initially. In this irrotational flow regime, $\buL$ is diagnosed from $\pse$ as $\buL = \pse - \bnabla \phi$ for a potential $\phi$ that is uniquely determined by the Poisson equation $\nabla^2 \phi = \bnabla \bcdot \pse$ and the boundary conditions satisfied by $\buL$. The CWCM thus reduces to the action equation \eqref{waveAct}, with $\bcU(\bx,\kappa)$ related linearly to $\act(\bx,\bk)$ via solution of the Poisson equation.


For potential vorticity, suppose that we have a passive scalar $b(\bx,z,t)$ satisfying wave-averaged material conservation,
\beq
\pt b + \buL\bcdot \grad b = \bcirc\per 
\label{ccons3}
\eeq
Potential vorticity is
\begin{align}
\Pi = \bzetaa \bcdot \grad b\com \label{PV1}
\end{align}
and satisfies 
\beq
\p_t \Pi + \buL\bcdot \grad \Pi =\diver \left( \bucirc \cross \grad b +  \bzetaa \bcirc \right) \per
\eeq


\section{Variational derivation} \label{varsec}

The conservation properties of the CWCM  can be traced to its variational formulation. We obtain this variational formulation from that of the rotating incompressible Euler equations.

\subsection{Rotating Euler equations}

Variational formulations of fluid models typically involve the flow map, $\bphi$, which sends labels $\ba$ identifying fluid particles to the time-$t$ positions of the particles, $\bx = \bphi(\ba,t)$ \citep[e.g.][]{salmon1988hamiltonian}. Incompressibility implies that the flow map preserves volume, that is, $| \partial \bphi / \partial \ba| = \mathrm{const}$. We can take the constant to be $1$ without loss of generality.
Flow map and velocity field are related via
\beq
\partial_t \bphi(\ba,t)=\bu(\bphi(\ba,t),t).
\eeq

Following \citet{Dewar1970}, \citet{bret71} and \citet{gjajaholm} \citep[see also][]{holm2023lagrangian}, we make minimal use of labels and flow map by adopting an Eulerian variational formulation. In this formulation -- an instance of the Euler--Poincar\'e framework \citep{holm1998euler} -- the flow map only appears implicitly through a constrained form of the variations of $\bu$ 
first obtained by \citet{newcomb1961} and \citet{bret70}. 

In Appendix \ref{app:vareuler}, we show that the rotating, incompressible Euler equations, with a free surface at $z = s(\bx,t)$, follow from Hamilton's principle $\delta \int \mathcal{L} \, \dd t  = 0$ applied to the Lagrangian
\beq
\mathcal{L}[\bu,s] = \iint \dd \bx \left( \int_{-d}^s \left( \tfrac{1}{2}|\bu|^2 + \bm{h} \bcdot \bu \right) \, \dd z - \tfrac{1}{2} g s^2 \right).
\label{lagrangian1}
\eeq
The Eulerian variations $\delta \bu$ and $\delta s$ are induced by  variations $\delta \bphi$ of the flow map. Introducing an Eulerian vector field $\bq(\bx,t)=(q_1,q_2,q_3)$ by 
\beq
\delta\bphi(\ba,t) = \bq(\bphi(\ba,t),t),
\label{qdef}
\eeq
leads to 
\begin{subequations} \label{constvar}
\begin{align}
\delta \bu &= \partial_t \bq + \bu \bcdot \nabla \bq - \bq \bcdot \nabla \bu,  \label{var1} \\
\delta s &= q_3 - q_1 \partial_x s - q_2 \partial_y s, \label{var2}
\end{align}
\end{subequations}
where  $\bq$ satisfies $\grad \bcdot \bq = 0$ and is otherwise arbitrary (see appendix \ref{app:vareuler}). Incompressibility is enforced by considering only volume preserving flow maps and hence divergence-free $\bu$ and $\bq$. No explicit constraint  enforcing $\diver \bu=0$ is therefore required in \eqref{lagrangian1}.

\subsection{Lagrangian mean framework}

Following the GLM framework, we introduce a wave--mean flow decomposition at the level of the flow map by writing
\beq
\bphi(\ba,t) = \bar {\bphi}(\ba,t) + \bxi(\bar{\bphi}(\ba,t),t), 
\label{decomp}
\eeq
where $\bar{\bphi}$ and $\bxi$ denote the mean flow and  perturbation maps. Below we introduce a linear-wave ansatz for $\bxi$  that captures the wavy displacements resulting from surface gravity waves; $\bar{\bphi}(\ba,t)$ captures the mean currents.
 
Differentiating \eqref{decomp} with respect to $t$ and using the mean position ${\bx_3}=(\bx,z)=\bar {\bphi}(\ba,t)$ as independent variable gives
\beq
\bu^{\bxi} = \buL + D_t \bxi.  
\label{fm11}
\eeq
Here, $\buL(\bx,t)$ defined by
\beq
\buL(\bar \bphi(\ba),t) = \partial_t \bar \bphi(\ba,t)
\eeq
is  the Lagrangian mean velocity,
\beq
D_t = \partial_t + \buL \bcdot \grad \com
\eeq
and 
\beq
\bu^{\bxi} (\bx_3,t)= \bu(\bx_3 + \bxi(\bx_3,t),t).
\eeq

The decomposition \eqref{decomp} involves a degree of arbitrariness. We take advantage of this to impose that the mean flow map preserves volume and leaves the free surface flat, that is,
\beq
\left| \partial \bar \bphi / \partial \ba \right| = 1 \quad \textrm{and} \quad \bar \varphi_3(a,b,c=0)=0. \label{contrast}
\eeq
In contrast with standard GLM, the assumptions in \eqref{contrast} imply a non-zero average for the perturbation map, $\overline{\bxi} \not=0$  \citep[see][]{vanneste2022}. Below in \eqref{ansatz}, however,  we use a linear approximation for $\bxi$ so that $\overline{\bxi}$ vanishes up to neglected terms. 


Volume preservation implies that $\buL$ is divergence free, i.e.\ \eqref{divuL}. With this, we can rewrite the Lagrangian \eqref{lagrangian1} using the mean position $(\bx,z)= \bar \bphi(\ba)$ as an independent variable. This results in 
\beq
\mathcal{L}[\buL,\bxi] = \iint  \dd \bx \left( \int_{-d}^{0} \tfrac{1}{2}|\buL + D_t \bxi|^2  + \bm{h}^{\bxi} \bcdot (\buL + D_t \bxi)  \, \dd z - \tfrac{1}{2} g \zeta^2 \right),
\label{lagrangian1p}
\eeq
where $\zeta = \xi_3(\bx,z=0)$ and $\bm{h}^{\bxi}(\bx_3)=\bm{h}(\bx_3 + \bxi(\bx_3))= \bh(\bx_3) + \half \bff \cross \bxi(\bx_3)$. 
(Without the assumption of volume preservation \eqref{contrast}, \eqref{lagrangian1p} would contain a Jacobian factor $\tilde \rho$,  as in the GLM formulation of \citet{AndrewsMcIntyre1978}.)




\subsection{Averaged Lagrangian}

Up to \eqref{lagrangian1p} no  approximations have been made. We now make a WKB ansatz for $\bxi$, assuming a scale separation between rapid phase variations and slow envelope modulations, and Whitham average. Following the prescription of \citet{whit65b,whit74}, the displacement field $\bxi$  is approximated by that of a linear wavetrain,  
\beq
\bxi \approx \Re  a(\bx,t) \hat \bxi(\bx,z,\bk)\ee^{\ii \theta(\bx,t)} \com 
\label{ansatz}
\eeq
where $a(\bx,t)$ is a slowly varying amplitude, $\theta(\bx,t)$ the phase, and $\bk=(k,l) = \bnabla_{\bx} \theta$. The vertical structure is dictated by 
\beq
\hat{\bxi}(\bx,z,\bk) = \frac{1}{\kappa \sinh(\kappa \ch{d(\bx)})} \left( \begin{array}{c} \ii k \cosh(\kappa(z+\ch{d(\bx)}) \\ \ii l \cosh(\kappa(z+\ch{d(\bx)}) \\ \kappa \sinh(\kappa(z+\ch{d(\bx)})) \end{array} \right),
\label{vstru}
\eeq
with $\kappa = |\bk| = |\grad_{\bx} \theta|$.
This structure is that  obtained for surface waves propagating in a locally uniform current and with locally uniform depth  \ch{as is usual with the WKB approximation}. \ch{We also neglect the effect of rotation on the waves, assuming $f \ll \sigma$.} 

With $\hat{\bxi}$ in  \eqref{vstru}, the displacement $\bxi(\bx,t)$ in \eqref{ansatz} is approximately incompressible and irrotational:
\beq
\grad \bcdot \bxi  \approx 0\com \qquad \text{and} \qquad \curl \bxi \approx 0\per
\label{WhitWKB}
\eeq
This approximation  is associated with Whitham's variational derivation of the WKB approximation: applying $\grad=(\bnablax,\p_z)$ to \eqref{ansatz} the derivatives act only on the phase $\theta(\bx,t)$ and on the $z$-dependence of $\hat{\bxi}$. For instance,
\beq
 D_t \bxi \approx \Re   a(\bx,t) \left( \hat \bxi(\bx,z,\bk)(-\ii \omega + \ii \buL\bcdot \bk)
+  \wL(\bx,z)\hat \bxi_z(\bx,z,\bk) \right) \ee^{\ii \theta(\bx,t)} \per  \label{secline}
 \eeq
We further neglect the term involving $\wL \hat \bxi_z$ in \eqref{secline} on the grounds that $\wL \ll \uL, \vL$. 

%
%

Introducing \eqref{ansatz} into the Lagrangian \eqref{lagrangian1p} and averaging over the phase $\theta$ leads to the averaged Lagrangian
\begin{align}
\overline{\mathcal{L}}[\buL,\theta,a] &= \iint \dd \bx \left( \int_{-d}^0 \left( \tfrac{1}{2}|\buL|^2 + \bm{h}(\bx) \bcdot \buL \right) \, \dd z   \right. \label{lagrangian2-} \\ 
& \qquad  \qquad + \left. \tfrac{1}{4}  \left(  \int^0_{-d}  \frac{\cosh(2 \kappa (z+d))\left(\omega - \bk \bcdot \buL \right)^2}{\sinh^2(\kappa d)} \, \dd z - g \right)|a|^2  \right), \nonumber
\end{align}
where $\omega = -\partial_t \theta$ and we neglect the wave contributions to the Coriolis term.

Using the definition of $Q(\bx,z,\ka)$ in  \eqref{Q}, we rewrite the second $z$-integral  in \eqref{lagrangian2-}  as
\beq
\int^0_{-d}  \frac{\cosh(2 \kappa (z+d))\left(\omega - \bk \bcdot \buL \right)^2}{\sinh^2(\kappa d)} \, \dd z = \frac{1}{\ka \tanh(\ka d)} \int^0_{-d} \left(\omega - \bk \bcdot \buL \right)^2 Q(\bx,z,\ka) \, \dd z\per 
\eeq
Justified by the smallness of the Doppler shift against the wave frequency, we make the approximation
\begin{subequations} \label{approx71}
\begin{align}
\int_{-d}^0 (\omega - \bk \bcdot \buL)^2 Q \, \dd z &= \omega^2 - 2 \omega \bk\bcdot \int^0_{-d}\!\! \buL Q \, \dd z +  \int^0_{-d}\!\! (\bk\bcdot \buL)^2 Q \, \dd z\com \label{approx70} \\ 
&\approx (\omega - \bk \bcdot \bcU)^2\com 
\end{align}
\end{subequations}
where  the Doppler velocity is 
\beq
\bcU =\int^0_{-d} \buL Q \, \dd z \com
\eeq
as in \eqref{Udop}. 
This manoeuvre simplifies \eqref{lagrangian2-} to
\beq
\overline{\mathcal{L}}[\buL,\theta,a] = \iint \dd \bx \left( \int_{-d}^0 \left( \tfrac{1}{2}|\buL|^2 + \bm{h}(\bx) \bcdot \buL \right) \, \dd z   + \tfrac{1}{4}  \left(   \frac{\left(\omega - \bk \bcdot \bcU \right)^2}{\kappa \tanh(\kappa d)} - g \right)|a|^2  \right).
\label{lagrangian2}
\eeq

The approximation in \eqref{approx71} is not necessary: one can proceed from the Lagrangian in  \eqref{lagrangian2-}. The resulting model has a more complicated dispersion relation  than \eqref{omIntro}, and a more complicated expression for the pseudomomentum than \eqref{pmom1}. Alternatively, one obtains a superficially simpler model by boldly dropping the final term in \eqref{approx70}. But this is one step too many: we prefer   the dispersion relation \eqref{omIntro} with a physically intuitive Doppler shift that preserves Galilean invariance.




The Lagrangian \eqref{lagrangian2} is the surface-wave analogue of the Lagrangian obtained by \citet{bret71}, \cite{gjajaholm} and \citet{salm16} for acoustic and gravity waves. Hamilton's principle can be applied using  variations $\delta \buL$ similar to \eqref{var1}, $\delta a$ and $\delta \theta$. This leads to the Craik--Leibovich system and to equations governing the phase $\theta(\bx,t)$ and amplitude $a(\bx,t)$ of a single wavetrain. We do not detail these equations here. Instead, we follow \citet{Dewar1970} to relax the assumption of a single wavetrain and consider a spectrum of linear waves.

\subsection{Variational form of the CWCM}

A spectrum of waves can be regarded as a superposition of independent wavetrains indexed by a two-dimensional (continuous) variable, $\balpha$ say (see \citeauthor{Dewar1970}'s (\citeyear{Dewar1970}) discussion of the `weak turbulence' assumption). Phase $\theta(\bx,\balpha,t)$ and amplitude $a(\bx,\balpha,t)$ then depend on $\balpha$, and the Lagrangian \eqref{lagrangian2} acquires an additional integration over $\balpha$ to become
\begin{align}
\overline{\mathcal{L}}[\buL,\theta,a] &= \iint \dd \bx \left( \int_{-d}^0 \left( \tfrac{1}{2}|\buL|^2 + \bm{h}(\bx) \bcdot \buL \right) \, \dd z   \right. \nonumber \\ 
& \qquad  \qquad + \left. \tfrac{1}{4}  \iint \left(\frac{\left(\omega - \bk \bcdot \bcU \right)^2}{\kappa \tanh(\kappa d)}- g \right) |a|^2 \, \dd \balpha \right) \per \label{lagrangian4}
\end{align}

The representation using $(\bx,\balpha)$ as independent variables and $(\theta,a)$ as dependent variables is inconvenient. It is preferable to use instead $(\bx,\bk)$ as independent variables and $(\omega,\act)$ as dependent variables, where
\beq
\act =    \frac{\sigma |a|^2}{2 \kappa \tanh(\kappa d)} \left|\frac{\partial\bk}{\partial  \balpha} \right|^{-1}.
\eeq
\ch{This specific choice ensures that $\act$ is dual to the frequency $\omega$ in the sense that $\act = \delta \overline{\mathcal{L}} / \delta \omega$ (see appendix \ref{app:varcoupled}). This identifies $\act$ as the wave action density in $(\bx,\bk)$-space \citep{Dewar1970}.} Making the change of variables in \eqref{lagrangian2} puts the averaged Lagrangian in its final form
\begin{align}
\mathcal{L}_{\textsc{cwc}}[\buL,\act,\omega] &= \iint \dd \bx \left( \int_{-d}^0 \left( \tfrac{1}{2}|\buL|^2 + \bm{h}(\bx) \bcdot \buL \right) \, \dd z   \right. \nonumber \\
& \qquad \qquad \qquad \left. + \, \tfrac{1}{2}  \iint \left(  \sigma^{-1} \left(\omega - \bk \bcdot \bcU\right)^2 - \sigma \right) \act \, \dd \bk \right).
\label{lagrangian3}
\end{align}
The Lagrangian $\mathcal{L}_{\textsc{cwc}}$ is the Lagrangian of the CWCM.
The term on the second line is a standard Whitham-averaged Lagrangian governing wave dynamics. Because of its dependence on the current variable $\buL$ via $\bcU$, this term contributes to the dynamics of both waves and currents. This is key to  consistency of the CWCM.

Hamilton's principle can now be applied to \eqref{lagrangian3}. The variations are
\beq
\delta \buL = \p_t \bq + \buL \bcdot \bnabla \bq - \bq \bcdot \bnabla \buL,  
\label{varul}
\eeq
with $\bq$ satisfying $\grad \bcdot \bq =0$ but otherwise arbitrary (in analogy with \eqref{var1}), $\delta \act$, and
\beq
\delta \omega = \partial_t r + \bnabla_{\bx} r \bcdot \bnabla_{\bk} \omega - \bnabla_{\bk} r \bcdot \bnabla_{\bx} \omega 
\label{varomega}
\eeq
with $r$ arbitrary. This leads to the Craik--Leibovich equation in the form \eqref{AML}, dispersion relation \eqref{omIntro} and  action transport equation \eqref{waveAct}. Details are given in Appendix \ref{app:varcoupled}. The form \eqref{varomega} of the variation of $\omega$, reminiscent of \eqref{varul}, emerges when changing variables from the mixed Eulerian-Lagrangian $\theta(\bx,\balpha)$ to the Eulerian $\omega(\bx,\bk)$ (see \citet{Dewar1970} and Appendix \ref{app:varomega}).


\section{Hasselmann's problem} \label{sec:hasselmann}

\cite{Hassel1970} showed that the rectified effect of surface waves is to drive inertial oscillations through `a vertical shear stress  induced by the rotation of the earth';  see also \cite{Ursell1950}. To exhibit this nonlinear interaction between surface waves and inertial oscillations, we consider a flat-bottom ($\bnablax d=0$), initially motionless ocean with no waves and no currents:
\beq
 \act (\bx,\bk) =\buL(\bx,z)=0 \qquad \text{at} \ \ t=0 \per
\eeq
Currents and waves are jointly forced by suddenly switching on horizontally uniform wind forcing. This $t>0$  forcing  is modelled with 
\beq
\actcirc = \alpha \big(\actst(\bk)  -\act(\bx,\bk)\big)
\label{simp}
\eeq
on the right-hand side of the action transport equation \eqref{waveAct}.
In \eqref{simp} $\actst(\bk)$ is an equilibrium wave spectrum  with  pseudomomentum 
\begin{subequations}
\begin{align}
\psest (z) &= \iint  Q(z,\kappa)  \actst(\bk) \bk \, \dd \bk   \\
&=\psst(z) \eye \per
\end{align}
\end{subequations}
The parameter  $\alpha$ in \eqref{simp} controls the rate at which the action spectrum adjusts to the equilibrium spectrum $\actst$. In Hasselmann's idealization, $\actst(\bk)$ and $\Om=\sig(\ka) + \bcU(\ka)\bcdot \bk$ do not depend on the horizontal coordinates $\bx=(x,y)$.  Thus there is  a  horizontally uniform  solution of  the action equation \eqref{waveAct}:
\beq
\act(\bk,t) = (1-\ee^{-\alpha t}) \actst(\bk)\per
\eeq
The  time-varying pseudomomentum is
\begin{align}
\pse(z,t) &= \big(1-\ee^{-\alpha t}\big) \psst(z) \eye \per
\label{unsteady}
\end{align}

Making the traditional approximation that $\bff = f \kay$ and assuming  a solution with the form $\buL = (\uL(z), \vL(z),0)$, we have
\beq
\curl (\buL- \pse)\cross  \buL = - \buL \bcdot \p_z (\buL - \pse) \, \kay \per
\label{vertForce}
\eeq
This vertical force is balanced by the vertical pressure gradient. 

With $\bucirc=0$, the horizontal components of $\buL$ evolve according to
\begin{subequations}
\begin{align}
\pt \buL + f \kay \cross \buL &= \pt \pse \label{boh?}  \\
&=  \alpha \ee^{-\alpha t} \psst \eye \per
\label{horizComp}
\end{align}
\end{subequations}
The right hand side of \eqref{boh?} is the inviscid forcing of currents resulting from  the `growth of swell' \citep{WagnerChini2021}.

\begin{figure}
  \centering
  \includegraphics[width=0.8\textwidth]{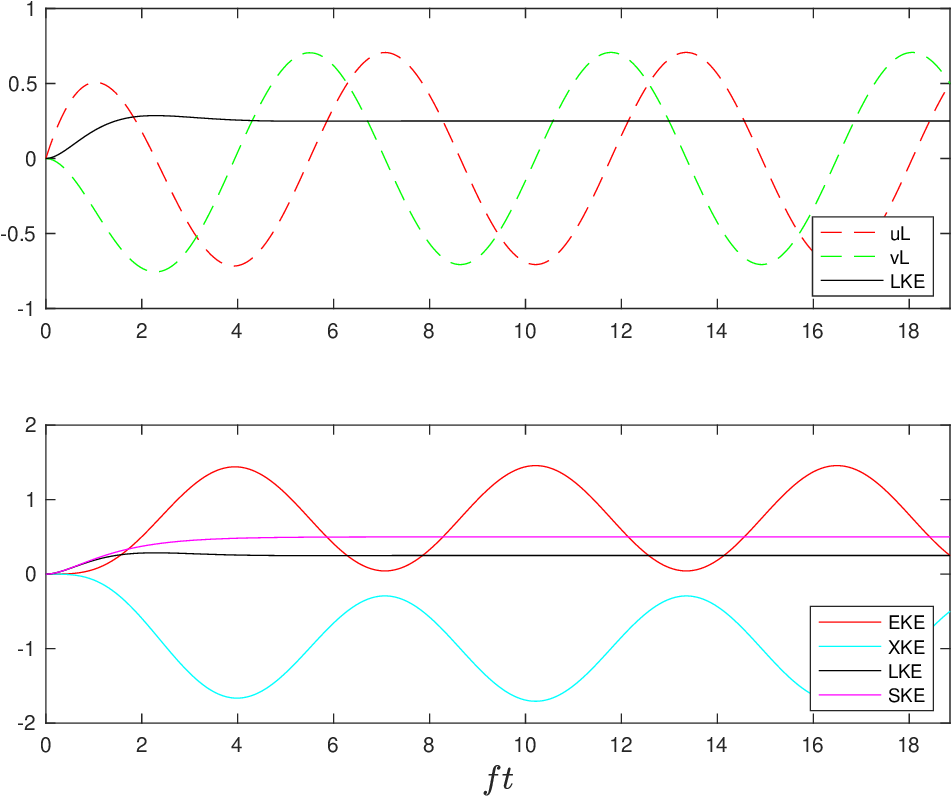}
  \caption{Solution of the Hasselmann problem with $\al=f$. Upper  panel: the dashed sinusoids  uL and vL  are the scaled Lagrangian velocities $\uL(z,t)/\psst(z)$ and $\vL(z,t)/\psst(z)$ and  LKE $= |\buL|^2/2 \psst^2$, is the corresponding scaled Lagrangian kinetic energy. Lower panel: the four kinetic energy densities defined in \eqref{KEz}, all scaled with $\psst(z)^2$. \ch{In this illustration $\alpha/f=1$.}}
  \label{figure2}
\end{figure}

The solution of \eqref{horizComp}, shown in the upper panel of figure \ref{figure2}, is
\begin{subequations}
\begin{align}
\uL+\ii \vL  &= \frac{\alpha \psst}{\alpha - \ii f} \big(\ee^{-\ii ft} - \ee^{-\alpha t}  \big)  \\
&\sim  \frac{\alpha \psst \ee^{-\ii ft}}{\alpha - \ii f}  \quad \text{as} \ \ t\to \infty \per
\label{IOsol}
\end{align}
\end{subequations}
The long-time Lagrangian  kinetic energy of the residual  inertial oscillation is therefore
\beq
\half |\buL|^2 = \half \frac{\alpha^2 \psst^2}{\alpha^2 + f^2}\per
\label{RIO}
\eeq
The total work done by the wind in establishing the long-time solution is the sum of the wave energy and the energy in the residual  inertial oscillation \eqref{RIO}:
\beq
E_{\mathrm{w}}+ E_{\mathrm{io}} = \iint \sigma(\ka) \actst(\bk)\, \dd \bk + \half \int_{-d}^0 \frac{\alpha^2 \psst^2(z)}{\alpha^2 + f^2}\, \dd z\per
\label{Ew+Eio}
\eeq
In terms of wave steepness $\epsilon$, $E_{\mathrm{w}}$  is order $\epsilon^2$ while $E_{\mathrm{io}}$  of order $\epsilon^4$.

Equation \eqref{Ew+Eio} shows that in Hasselmann's scenario the excitation of inertial waves does not deplete wave energy but is instead achieved through an increase of the wind work. The energy balance \eqref{Econs5} is consistent with this conclusion: combining  the expression for energy production in \eqref{energ31} with $\Omega = \sigma + \bk \bcdot \bcU$, \eqref{Econs5}  reduces to
\beq
\pt E   = \underbrace{\iint \sigma \actcirc \dd \bk}_{\pt E_\mathrm{w}} +\underbrace{\iint \bk \bcdot \bcU \actcirc \dd \bk}_{\pt  E_{\mathrm{io}} }\per
\label{Hassel17}
\eeq
The final term in \eqref{Hassel17} is the production by wind of near-inertial energy, $E_{\mathrm{io}}$  in \eqref{Ew+Eio}.

The assumption of horizontal uniformity ensures that  $\bnablak \, \Om \bcdot \bnablax \act - \bnablax\, \Om \bcdot \bnablak  \, \act = 0$ in \eqref{waveAct}, i.e.\ the inertial current in \eqref{RIO} has no effect on $\act(\bx,\bk,t)$. It  is possible, however, that the inertial oscillation  is linearly unstable, and saturation of this instability might lead to turbulent currents and modification of $\act$ by currents. This possibility  is strongly indicated  by the numerical solutions  of \cite{WagnerChini2021} in which  the  pseudomomentum $\pse(z,t)$ is prescribed. In fact  prescription of $\pse(z,t)$ is recovered as  the special case $\alpha \to \infty$ of the relaxation-to-$\actst$  model in \eqref{simp}.

The `Lagrangian kinetic energy density'  is partitioned by the decomposition $\buL = \buE+ \buS$ (with $\buS \approx \pse$) into
\beq
\underbrace{\half |\buL|^2}_{\text{LKE}} = \underbrace{\half |\buE|^2}_{\text{EKE}} + \underbrace{\buS\bcdot \buE}_{\text{XKE}}+ \underbrace{\half |\buS|^2}_{\text{SKE}}\per
\label{KEz}
\eeq
Above LKE is the Lagrangian kinetic energy (which produces $E_{\mathrm{io}}$ when integrated in $z$), EKE is the Eulerian kinetic energy, XKE is the kinetic energy in the `cross term' $\buE\bcdot \buS$ and  SKE is the Stokes kinetic energy.
 The lower panel of figure \ref{figure2} shows these four kinetic energy densities in  the  $\alpha/f=1$ solution. LKE is   smaller in magnitude than the three terms on the right of \eqref{KEz} and, unlike EKE and XKE, has simple monotonic evolution. The same conclusions are drawn by \citet{czeschel2023energy}. 

In the long-time limit  $\ee^{-\al t}\to 0$, the decomposition in \eqref{KEz} is
\beq
\underbrace{\half \frac{\alpha^2 \psst^2}{\alpha^2 + f^2}}_{\text{LKE}} =  \underbrace{\half |\buE|^2 + \buE\bcdot \buS}_{\text{EKE + XKE}}  + \underbrace{\half \psst^2}_{\text{SKE}}\per \label{ELmess}
\eeq
 EKE and XKE both oscillate at the inertial frequency, but their sum is steady and negative:
\beq
\text{EKE + XKE} = - \half \frac{f^2 \psst^2}{\al^2+ f^2} \leq 0\per
\eeq
In the long-time limit  XKE is negative and greater in magnitude than EKE for all values of $\al/f$. 

When viewed in terms of the monotonically increasing  LKE in figure \ref{figure2},  the  energetics of Hasselmann's problem is straightforward: wave forcing via $\pse(z,t)$ is steadily increasing the LKE of the currents.   But when viewed in terms  of EKE and XKE  this simple result is obscured and the problem seems complicated and confusing. This Lagrangian simplicity is because $|\buL|^2/2$ is the current kinetic energy that participates in the complete and consistent energy conservation law in \eqref{Econs3} and \eqref{Econs5}.


The Doppler velocity takes the form
\begin{align}
\Uc + \ii \Vc &= \int_{-d}^0 Q(\ka,z) (\uL+\ii \vL)\, \dd z\nonumber \\
&= \frac{\al}{\al - \ii f} \big(\ee^{-\ii ft} - \ee^{-\al t} \big) \int \Lambda(\ka,\ka') \actst(\bk') \bk' \, \dd \bk', \label{UU}
\end{align}
where
\beq
\Lambda(\ka,\ka') = \int_{-d}^0\!\!\!Q(\ka,z)Q(\ka',z)  \,  \dd z  = \frac{2 \ka\ka'}{\ka+ \ka'} \frac{\ka \coth (2 \ka'd) - \ka' \coth (2 \ka d)}{\ka - \ka'}.
\eeq
The general argument of \S\ref{energyCons} shows that introducing \eqref{UU} into the right-hand side of \eqref{Hassel17} recovers $\partial_t E_\mathrm{io}$.

\section{Discussion}

CWCM, the model of two-way coupling between surface waves and ocean currents introduced in \S\ref{dcore}, is perhaps the simplest  that conserves momentum and energy. This simplicity is achieved in part by adopting a wave--mean separation that differs from standard GLM to ensure (i) that the mean current is exactly divergence free, and (ii) that the mean free surface is flat. Property (ii) amounts to restricting mean configurations to those that minimise gravitational potential energy. Such a restriction is standard in the treatment of dynamical systems under a strong constraining potential \citep[e.g.][]{takens2006motion}. It has the benefit of filtering surface waves from the mean equations. Property (ii) does not preclude capturing mean deformations in the free-surface elevation: in principle, these mean deformations  are represented through non-zero mean perturbations, $\overline{\bxi} \not= 0$. Our formulation, however, ignores these by following Whitham and using the linear approximation \eqref{vstru} for $\bxi$.  

The simplicity of the CWCM also stems from the variational approach used for its derivation. Compared with formal asymptotic expansions carried out at the level of the dynamical equations, the variational approach has the advantage of preservation of symmetry-related conservation and ease of implementation. (Exploratory calculations indicate that systematic asymptotics to fourth order in  wave steepness, as required to capture  terms such as $|\buS|^2/2$ in the energy equation, is exceedingly complicated.) Against this, an inconvenience is the difficulty in ascertaining the asymptotic accuracy of the variational model. But for a problem as complex as the two-way interactions between surface gravity waves and currents, involving multiple small parameters and  regimes of physical interest, we believe that the variational formulation has overwhelming advantages.
%
%

The CWCM in \S\ref{dcore}, with conservation laws on display in \S\ref{seccons},  follows from the application of Hamilton's principle to the Lagrangian ${\mathcal{L}}_{\textsc{cwc}}$ in \eqref{lagrangian3}.   The primary approximation  used in the construction $\mathcal{L}_{\textsc{cwc}}$ is Whitham's linear-wave approximation in \eqref{ansatz} and \eqref{vstru}. We then make a few secondary approximations, such as  simplification of the Doppler shift as in \eqref{approx71}. These secondary approximations can be avoided at the expense of  a more complicated model. For example, retaining $\wL\hat \bxi_z$ in \eqref{secline} results in  a three-dimensional pseudomomentum. Presumably this more complicated model is also a more accurate representation of the two-way interaction between currents and waves. 

%
%
%

Controversial issues such as the divergence and vertical component of $\pse \approx \buS$ or whether the current kinetic energy involves $\buE$ or $\buL$ do not arise in our formulation. Only the Lagrangian mean velocity $\buL$ and pseudomomentum $\pse$ play a dynamical role. The pseudomomentum $\pse$ appears as the variational derivative of the wave Lagrangian with respect to $\buL$, a  vector directly related to wave action, rather than as an approximation to the general three-dimensional form of \citet{AndrewsMcIntyre1978}. The primacy of $\buL$ and $\pse$ over $\buE$ and $\buS$ is an unavoidable feature of variational formulations of coupled wave--mean models. This aspect of the CWCM will remain if some of the approximations we have made are relaxed. 

The CWCM provides a sound basis for revisiting problems that have previously been examined with prescribed $\buS$ or $\pse$.  
The generation of inertial oscillations by a uniform field of surface waves considered in \S\ref{sec:hasselmann} is the simplest such problem. Because of the assumption of horizontal homogeneity, the dynamics of action is not affected the mean current that is generated. The main benefit of the CWCM is then to provide a consistent picture of the energy exchanges between wind, waves and currents. It will be interesting to consider problems in which currents and wave action vary in the horizontal so that the two-way interactions come into play. These problems include instabilities such as the Craik--Leibovich instability responsible for the formation of Langmuir cells. Stratification effects can be included straightforwardly by treating buoyancy as a passive scalar, as in  \eqref{ccons3}, and adding the buoyancy force in the momentum equation.

Applications of the Craik--Leibovich system to realistic ocean conditions often involve turbulence. The exchanges between waves and currents are then complicated by the separation of the currents into mean and turbulent components and the need to parameterise turbulent fluxes \citep[e.g.][]{SuzukiFoxKemper,czeschel2023energy}. Work along these lines should use the CWCM or a similar model that recognises the importance of the Lagrangian kinetic energy $|\buL|^2/2$.

\backsection[Acknowledgments]{We thank Onno Bokhove, Lars Czeschel and Greg Wagner for useful discussions and the anonymous referees for their valuable comments.}

\backsection[Funding]{JV is  supported by the UK Natural Environment Research Council (grant NE/W002876/1). WRY  is supported by the National Science Foundation award 2048583.}

\backsection[Declaration of interests]{The authors report no conflict of interest.}


\backsection[Author ORCID]{J. Vanneste, https://orcid.org/0000-0002-0319-589X,  W.R. Young,  https://orcid.org/0000-0002-1842-3197.}

\appendix

\section{ The  integral  involving $\pal \Om$ in \eqref{momCons4} } \label{vardepth}

\ch{The most difficult term in \eqref{momCons4}  is 
\begin{align}
\iint \act \pal \Om \, \dd \bk &= \iint \act \pal \Big[ \sig\big(d(\bx),\bk\big)+ \int_{-d(\bx)}^0 \!\!\!\!\!\!\! Q\big(d(\bx),z,\ka \big) \bk\bcdot \buL(\bx,z)  \, \dd z \Big] \, \dd \bk\per \label{diff3}
\end{align}
On the right of \eqref{diff3} we have written the Doppler shifted frequency $\Om(\bx,\bk)$ expansively and indicated that the $\bx$-dependence of $\sigma$ and $Q$ is  exclusively through the depth $d(\bx)$. Notice that the derivative $\pal$ acts on $d(\bx)$ and also on the separate $\bx$-dependence of $\buL(\bx,z)$. We have
\beq
\iint \act \pal \Om \, \dd \bk = \iint \act  \int_{-d}^0 \!  Q(d,z,\ka) \bk\bcdot \pal \buL\, \dd z \, \dd \bk + \iint \act  \frac{\p \Om}{\p d} \, \dd \bk\, \pal d \per \label{dHell1}
\eeq
Most of the complication is in
\beq
 \frac{\p \Om}{\p d} =  \frac{\p \sigma}{\p d} + \bk\bcdot  \!\int_{-d}^0  \frac{\p Q}{\p d} \buL(\bx,z) \, \dd z + \bk\bcdot  [Q\buL]_{-d}\per \label{dHell3}
\eeq
Above $[]_{-d}$ indicates evaluation at $z=-d(\bx)$ i.e.
$ [Q\buL]_{-d} = Q(d,-d,\ka) \buL(\bx,-d)$.
Switching the order of the $z$ and $\bk$-integrals in the first term on the right of \eqref{dHell1} produces
\beq
\iint \act \pal \Om \, \dd \bk  = \int^0_{-d} \!\!\! \ps_\bet \pal \uL_\bet \, \dd z +  \iint \act  \frac{\p \Om}{\p d} \, \dd \bk\, \pal d \com \label{dHell5}
\eeq
where we have used the definition of the pseudomomentum in \eqref{pmom1}. 
}

\ch{There is an interesting alternative expression for $\p\Om/\p d$ in \eqref{dHell3}.
We first establish a useful identity:   $\p_\alpha$ of the $Q(\bx,z,\ka)$ normalization integral \eqref{Qnorm} produces
\beq
\int_{-d}^0 \!\! \pal Q \, \dd z + [Q]_{-d} \pal d = 0\per
\label{lem3}
\eeq
Substituting  
\beq
\pal Q =\frac{\p Q}{\p d} \, \pal d
\eeq
 into \eqref{lem3} and cancelling  $\pal d$ produces the identity
\beq
\int_{-d}^0 \frac{\p Q}{\p d}  \dd z + [Q]_{-d}  = 0\per
\label{lem7}
\eeq
Using \eqref{lem7}  to eliminate $[Q]_{-d}$ from \eqref{dHell3}
\beq
 \frac{\p \Om}{\p d} =  \frac{\p \sigma}{\p d} + \bk\bcdot  \!\int_{-d}^0  \frac{\p Q}{\p d} \big(\buL - [\buL]_{-d}\big)\, \dd z \per \label{dHell7}
\eeq
In the shallow-water limit, with $\buL$ independent of $z$, only $\p\sig/\p d$ remains.} 

\section{Variational derivation of the rotating Euler equation} \label{app:vareuler}

We derive the rotating incompressible  Euler equations from Hamilton's principle applied to the Lagrangian \eqref{lagrangian1} with constrained variations \eqref{constvar}.
We start by deriving the constrained variations \eqref{constvar} and \eqref{varomega} following \citet{newcomb1961}, \citet{bret70} and \citet{Dewar1970}.

\subsection{Variations $\delta u$ and $\delta s$}

The variations $\delta \bu(\bx,t)$ of the velocity field are induced by variations $\delta \bphi(\ba,t)$ of the flow map. 
Taking the time derivative of \eqref{qdef} gives
\begin{align}
\pt \delta  \bphi(\ba,t) &= \pt \bq (\bphi(\ba,t),t) + \pt \bphi(\ba,t) \bcdot \bnabla \bq(\bphi(\ba,t),t) \nonumber\\
&= \pt \bq (\bphi(\ba,t),t) + \bu(\bphi(\ba,t),t) \bcdot \bnabla \bq(\bphi(\ba,t),t). \label{mixed1}
\end{align}
Another expression for the left-hand side is obtained by taking variations of $\pt \bphi(\ba,t)=\bu(\bphi(\ba,t),t)$ and using the equality $\delta \partial_t = \pt \delta$ of mixed derivatives  to find
\begin{align}
 \pt \delta  \bphi(\ba,t) &= \delta \bu(\bphi(\ba,t),t) + \delta \bphi(\ba,t) \bcdot \bnabla \bu(\bphi(\ba,t),t) \nonumber \\
&= \delta \bu(\bphi(\ba,t),t) + \bq(\bphi(\ba,t),t) \bcdot \bnabla \bu(\bphi(\ba,t),t). \label{mixed2}
\end{align}
Equating \eqref{mixed1} and \eqref{mixed2} gives \eqref{var1}.

The form \eqref{var2} of $\delta s$ is obtained by relating free surface elevation and flow map via
\beq
s(\varphi_1(a,b,0),\varphi_2(a,b,0),t) = \varphi_3(a,b,0,t),
\eeq
where $\ba=(a,b,0)$ parameterises the free surface in label space. Taking variations gives
\beq
\delta s(\bx,t) + \delta \varphi_1(a,b,0,t) \partial_x s(\bx,t) + \delta \varphi_2(a,b,0,t) \partial_y s(\bx,t)  =  \delta \varphi_3(a,b,0,t),
\eeq
that is,
\beq
\delta s(\bx,t) +  q_1(\bx,t) \partial_x s(\bx,t) +  q_2(\bx,t) \partial_y s(\bx,t) = q_3(\bx,t),
\eeq
where $\bx=(\varphi_1(a,b,0,t),\varphi_2(a,b,0,t))$. Rearranging gives \eqref{var2}. The assumption of volume preserving flow map implies that $\bnabla \bcdot \bq = 0$.

\subsection{Hamilton's principle applied to \eqref{lagrangian1}}

Hamilton's principle states that
\beq
\delta \int  \mathcal{L} \, \dd t = \delta \int \dd t  \iint \dd \bx \left( \int_{-d}^s \bu^\mathrm{a} \bcdot \delta \bu \, \dd z + \left( \left[ \tfrac{1}{2} |\bu|^2 + \bm{h} \bcdot \bu \right]_{s}  - g s \right)  \delta s \right) = 0, 
\label{hp}
\eeq 
where $\bu^\mathrm{a} =   \bu + \bm{h}$ and   $[\,\cdot\,]_s$ denotes evaluation at $z=s$. We use \eqref{var1} and manipulate the $z$ integral:
\begin{align}
\int_{-d}^s \bu^\mathrm{a} \bcdot \delta \bu \, \dd z &= \int_{-d}^s \bu^\mathrm{a} \bcdot (\pt \bq + \bu \bcdot \bnabla \bq - \bq \bcdot \bnabla \bu) \, \dd z\com \nonumber \\
&= \int_{-d}^s \Big( \pt (\bu^\mathrm{a} \bcdot \bq) + \bnabla \bcdot (\bu^\mathrm{a} \bcdot \bq \, \bu)  \nonumber \\
& \qquad - (\pt \bu + \bu \bcdot \bnabla \bu^\mathrm{a} + \bnabla \bu \bcdot \bu^\mathrm{a} ) \bcdot \bq  \Big) \, \dd z\per
\label{c2}
\end{align}
In $\bnabla \bu \bcdot \bu^\mathrm{a}$, the contraction is between $\bu$ and $\bu^\mathrm{a}$, \ch{i.e., in coordinates, $(\bnabla \bu \bcdot \bu^\mathrm{a})_i = (\p_i u_j) u^\mathrm{a}_j$. We repeatedly use this convention below.}
Integrating \eqref{c2} with respect to $\bx$ and using  the Reynolds transport theorem gives
\begin{align}
\iint \dd \bx &  \int_{-d}^s \bu^\mathrm{a} \bcdot \delta \bu \, \dd z = \dfrac{\dd}{\dd t} \iint \dd \bx  \int_{-d}^s \bu^\mathrm{a} \bcdot \bq \, \dd z  \nonumber  \\
& - \iint \dd \bx  \int_{-d}^s (\partial_t \bu + \bu \bcdot \bnabla \bu^\mathrm{a} + \bnabla \bu \bcdot \bu^\mathrm{a} ) \bcdot \bq  \Big) \, \dd z. \label{someeqn}
\end{align}
Here we impose the kinematic boundary condition 
$\partial_t s + u \partial_x s + v \partial_y s = w$ at $z=s(\bx,t)$ and the no-normal flow condition $w + \bu \bcdot \bnabla_{\bx}  d = 0$ at $z = -d(\bx)$.

The exact time derivative on the first line of \eqref{someeqn} does not contribute to \eqref{hp} because $\bq$ is assumed to vanish at the endpoints of the time integral. The second line makes a contribution that vanishes for all divergence-free $\bq$, as required, provided that
\beq
\partial_t \bu + \bu \bcdot \bnabla \bu^\mathrm{a} + \bnabla \bu \bcdot \bu^\mathrm{a}  = - \grad \pi
\label{euler1}
\eeq
for some scalar field $\pi$. (This follows from  Helmholtz decomposing $\partial_t \bu + \bu \bcdot \bnabla \bu^\mathrm{a} + \bnabla \bu \bcdot \bu^\mathrm{a}$ and observing that the vanishing of \eqref{someeqn} implies that the rotational part is zero.) Using vector calculus identities, \eqref{euler1} can be rewritten as
\beq
\partial_t \bu + (\bnabla \times \bu^\mathrm{a}) \times \bu = - \grad ( \pi + \bu^\mathrm{a} \bcdot \bu),
\label{euler2}
\eeq
then as
\beq
\partial_t \bu  + \bu \bcdot \bnabla \bu + \bm{f} \times \bu = - \grad p,
\label{euler3}
\eeq
where $p=\pi + \tfrac{1}{2} |\bu|^2 +\bm{h} \bcdot \bu$. This is the familiar form of the rotating Euler equation. 

With  \eqref{euler1} and using $\bnabla \bcdot \bq =0$, the second line in \eqref{someeqn} becomes
\beq
\iint \dd \bx  \int_{-d}^s \bnabla \bcdot ( \pi \bq)  \, \dd z = \iint \dd \bx \,  [\pi \,  \bm{n} \bcdot \bq]_s  = \, \iint \dd \bx \,  \left[ \pi \right]_s \delta s,
\eeq
where $\bm{n}=(-\nabla_{\bx} s,1)$ and we use the form \eqref{var2} of $\delta s$ and that $\bq$ is tangent to the bottom boundary. 
This reduces Hamilton's principle \eqref{hp} to  surface terms:
\beq
\delta \int  \mathcal{L} \, \dd t = \delta \int \dd t  \iint \dd \bx  \left( \left[ \pi + \tfrac{1}{2} |\bu|^2 + \bm{h} \bcdot \bu \right]_{s}  - g s \right)  \delta s.
\eeq
Imposing that these vanish leads to the dynamic boundary condition
\beq
p = g s.
\eeq

\section{Variational derivation of the CWCM} \label{app:varcoupled}

We derive the coupled model governed by \eqref{waveAct}--\eqref{rigidLid} from Hamilton's principle applied to the averaged Lagrangian ${\mathcal{L}}_\textsc{cwc}$ in \eqref{lagrangian2} with variations \eqref{varul} and \eqref{varomega}. The variations $\delta \act$ are unconstrained, hence Hamilton's principle implies that
\beq
\frac{\delta \overline{\mathcal{L}}}{\delta \act} = \sigma^{-1} ( \omega - \bk \bcdot \bcU)^2 - \sigma = 0. 
\eeq
The dispersion relation $\omega = \sigma + \bk \bcdot \bcU= \Omega(\bx,\bk)$ follows. Using this, we compute
\begin{align}
\frac{\delta \overline{\mathcal{L}}}{\delta \buL} &= \buL + \bm{h} - \iint  Q(\bx,z,\kappa)  \act(\bx,\bk) \bm{k} \, \dd \bk \nonumber \\
&= \buL + \bm{h} - \pse  =  \bupsa 
\end{align}
and
\beq
\frac{\delta \overline{\mathcal{L}}}{\delta \omega} = \act. 
\label{dual}
\eeq
With these results, Hamilton's principle reduces to
\beq
\delta \int  \mathcal{L} \, \dd t = \delta \int \dd t  \iint \dd \bx \left( \int_{-d}^0 \bupsa \bcdot \delta \buL \, \dd z + \iint \act \delta \omega \, \dd \bk \right) = 0. 
\label{hp2}
\eeq
The variations $\delta \buL$ and $\delta \omega$ are independent and can be treated separately. 

For $\delta \buL$, \eqref{varul} gives
\begin{align}
\iint \dd \bx  \int_{-d}^0 \bupsa \bcdot \delta \buL \, \dd z &= \iint \dd \bx  \int_{-d}^0 \bupsa \bcdot (\partial_t \bq + \buL \bcdot \bnabla \bq - \bq \bcdot \bnabla \buL) \, \dd z \nonumber \\
&= \iint \dd \bx  \int_{-d}^0  \partial_t (\bupsa \bcdot \bq) + \bnabla \bcdot (\bupsa \bcdot \bq \, \buL)  \label{c3} \\
& \qquad  \qquad \qquad - (\partial_t \bupsa  + \buL \bcdot \bnabla \bupsa + \bnabla \buL \bcdot \bupsa  ) \bcdot \bq   \, \dd z.
\nonumber
\end{align}
The terms on the first line make no contributions to \eqref{hp2}, bar boundary terms that vanish provided that the boundary conditions \eqref{rigidLid} and \eqref{bottombc} hold. The terms on the second line vanish for all divergence-free $\bq$ provided that
\beq
\partial_t \bupsa  + \buL \bcdot \bnabla \bupsa + \bnabla \buL \bcdot \bupsa = - \bnabla \hat \pi
\label{appCL}
\eeq
for some scalar $\hat \pi$. We recognise \eqref{appCL} as version \eqref{AML} of the Craik--Leibovich equation in vector form and with $\bucirc=0$.

For $\delta \omega$, \eqref{varomega} gives
\begin{align}
\iint \dd \bx \iint \act \delta \omega \, \dd \bk &= \iint \dd \bx \iint \act (\partial_t r + \bnabla_{\bx} r \bcdot \bnabla_{\bk} \omega - \bnabla_{\bk}r  \bcdot \bnabla_{\bx} \omega) \, \dd \bk \nonumber \\
&=\iint \dd \bx \iint \Big(\partial_t (\act r) +  \bnabla_{\bx} \bcdot (\act r  \bnabla_{\bk} \omega) - \bnabla_{\bk} \bcdot (\act r   \bnabla_{\bx} \omega) \nonumber \\
& \qquad \qquad - (\partial_t \act +  \bnabla_{\bk} \omega \bcdot \bnabla_{\bx} \act - \bnabla_{\bx} \omega \bcdot \bnabla_{\bk}\act) r \Big) \, \dd \bk
\end{align} 
Again, the terms in the first line are total derivatives and make no contribution to \eqref{hp2}. The vanishing of the remaining terms for arbitrary $r$ implies the wave action transport equation \eqref{waveAct} with $\actcirc=0$.

\section{Constrained variation $\delta \omega$} \label{app:varomega}

The variation $\delta \omega$ in \eqref{varomega}  arises from unconstrained the variation $\delta \theta(\bx,\balpha,t)$ via the change of dependent and independent variables
\beq
(\bx,\balpha,\theta(\bx,\balpha,t)) \mapsto (\bx,\bk,\omega(\bx,\bk,t))
\eeq
with 
\beq
\bk = \bK(\bx,\balpha,t) \defn \bnabla_{\bx} \theta(\bx,\balpha,t) \quad \textrm{and} \quad \omega(\bx,\bK(\bx,\balpha,t),t) = - \partial_t \theta(\bx,\balpha,t). 
\label{komthe}
\eeq
We define the function $r(\bx,\bk,t)$ by
\beq
r(\bx,\bK(\bx,\balpha,t),t) = - \delta \theta(\bx,\balpha,t).
\label{r}
\eeq
Taking the variation at fixed $\balpha$ of the second equation in \eqref{komthe} and using \eqref{r} gives
\beq
\delta \omega = \partial_t r + \partial_t \bK \bcdot \bnabla_{\bk} r - \delta \bK \bcdot \bnabla_{\bk} \omega.
\label{deltaomega}
\eeq
On the other hand, the variation and time derivative of the first of \eqref{komthe} give
\begin{subequations} \label{Kr}
\begin{align}
\delta \bK &= - \bnabla_{\bx} r - \bnabla_{\bx} \bK \bcdot \bnabla_{\bk} r, \\
\partial_t \bK &= - \bnabla_{\bx} \omega - \bnabla_{\bx} \bK \bcdot \bnabla_{\bk} \omega,
\end{align}
\end{subequations}
where the contractions are between $\bK$ and $\bnabla_{\bk}$.
Introducing \eqref{Kr} into \eqref{deltaomega} and using that $\bnabla_{\bx} \bK = \bnabla_{\bx} \bnabla_{\bx} \theta$ is a symmetric tensor leads to expression \eqref{varomega} for $\delta \omega$.

\bibliographystyle{jfm}
\bibliography{boring_rev.bib}
\end{document}